\documentclass[twocolumn]{pasj01}

\usepackage{mathpazo}
\usepackage[T1]{fontenc} 
\usepackage{lineno}
\newcommand\Ka{K{\ensuremath{\alpha}}~}

\Received{2022 November 29}
\Accepted{2023 February 28}
\Published{$\langle$publication date$\rangle$}

\begin{document}

\title{On the Zeeman Effect in Magnetically-Arrested Disks}
\author{Yoshiyuki \textsc{Inoue}\altaffilmark{1,2,3}}
\altaffiltext{1}{Department of Earth and Space Science, Graduate School of Science, Osaka University, Toyonaka, Osaka 560-0043, Japan}
\altaffiltext{2}{Interdisciplinary Theoretical \& Mathematical Science Program (iTHEMS), RIKEN, 2-1 Hirosawa, Saitama 351-0198, Japan}
\altaffiltext{3}{Kavli Institute for the Physics and Mathematics of the Universe (WPI), UTIAS, The University of Tokyo, Kashiwa, Chiba 277-8583, Japan}
\email{yinoue@astro-osaka.jp}


\maketitle

\begin{abstract}
Magnetically arrested disk (MAD) has been argued as the key accretion phase to realize the formation of relativistic jets. However, due to the lack of magnetic field measurements of accreting systems, MAD has not been observationally confirmed yet. Here we propose that a strong magnetic field accompanied by MAD would induce the Zeeman splitting of relativistically broadened Fe~\Ka fluorescence lines in X-ray binaries and active galactic nuclei, where we consider a two-phase medium in the inner accretion disk, magnetically dominated hot corona and cold reflector. Such a geometrical configuration is suggested from X-ray observations and recently confirmed by numerical simulations. Although turbulence in accretion flows would broaden the split lines, future X-ray high-energy resolution satellites, {\it XRISM} and {\it Athena}, would be capable of seeing the Zeeman effect on the Fe lines in X-ray binaries in the case with the MAD configuration. The signature of the Zeeman split lines would provide observational evidence for MAD.
\end{abstract}

\section{Introduction}
\label{sec:intro}

It has been more than a century since astrophysical jets were first observed \citep{Curtis1918PLicO..13....9C}. Astrophysical jets are collimated relativistic magnetized plasma outflows launched from compact accreting objects. They are found in stellar-mass black holes (BH; \cite{Fender2004MNRAS.355.1105F}) and supermassive BHs \citep{Blandford2019ARA&A..57..467B}. Powers of some relativistic jets can exceed the Eddington limit of BHs, requiring highly efficient energy conversion processes from accretion to outflows. However, the formation mechanism of powerful relativistic jets has yet to be answered.

Theoretically, the Blandford-Znajek (BZ) mechanism \citep{Blandford1977MNRAS.179..433B} is believed as the plausible explanation for the jet launch. In the BZ mechanism, the jet power is extracted by the rotation of BHs with the support of the magnetic fields threading the central BH. General relativistic magnetohydrodynamic (GRMHD) simulations confirm this process as a plausible and efficient jet power extraction mechanism (see, e.g., \cite{Komissarov2007MNRAS.380...51K, Tchekhovskoy2010ApJ...711...50T, Tchekhovskoy2011MNRAS.418L..79T, McKinney2012MNRAS.423.3083M, Takahashi2016ApJ...826...23T, Avara2016MNRAS.462..636A, Liska2022ApJ...935L...1L}). 

The BZ mechanism has two key parameters: the spin parameter of the BH and the large-scale magnetic field threading in the BH. Numerical simulations suggest that strong magnetic fields are required to realize observed powerful jets (e.g., \cite{McKinney2012MNRAS.423.3083M}). Such large magnetic flux accumulation is expected to appear in the magnetically arrested disk (MAD) scenario (\cite{Narayan2003PASJ...55L..69N, Igumenshchev2003ApJ...592.1042I}, see also \cite{Bisnovatyi-Kogan1974Ap&SS..28...45B, Bisnovatyi-Kogan1976Ap&SS..42..401B}), where magnetic field dominates the dynamics of the inner disk. Although the event horizon telescope (EHT) has resolved inner accretion disks of M~87 and Sgr~A* with unprecedented spatial resolutions, it is still not conclusive whether their accretion processes are dominated by MAD or other processes \citep{EHT2021ApJ...910L..13E, EHT2022ApJ...930L..16E, Blandford2022MNRAS.514.5141B}. Therefore, observational evidence of MAD in accretion systems is still lacking.

Here, the presence of a magnetic field induces spectral line splitting by the Zeeman effect due to the interaction of the magnetic dipole moment of an electron with the magnetic field. The Zeeman effect has been applied to measure magnetic fields of various astrophysical systems such as sunspots \citep{Hale1908ApJ....28..315H}, active stars \citep{Donati1997MNRAS.291..658D}, molecular clouds \citep{Nakamura2019PASJ...71..117N}, and an outer maser disk of an AGN \citep{Modjaz2005ApJ...626..104M}. Detectability of the Zeeman effect in the X-raying accreting neutron stars has also been argued in the literature \citep{Sarazin1977ApJ...216L..67S, Loeb2003PhRvL..91g1103L}.

BH-powered X-ray binaries (XRBs) and active galactic nuclei (AGNs) ubiquitously have the Fe~\Ka fluorescence line in their X-ray spectra. Broaden Fe~\Ka lines imply the location of the cold reflecting medium near the BHs and have been used for the BH spin measurements \citep{Reynolds2021ARA&A..59..117R}. Here, hot plasma, namely coronae, should also exist in the vicinity of BHs to reproduce X-ray continuum spectra of accreting BHs. Geometrical configurations of hot coronae and cold reflectors have been debated in literature (see e.g., \cite{Done2007A&ARv..15....1D, Meyer-Hofmeister2011A&A...527A.127M}). Recently, by performing two temperature GR-radiation-MHD simulations, \citet{Liska2022ApJ...935L...1L} demonstrated that a geometrically thin accretion disk transitions into a two-phase medium of cold gas clumps and a hot, magnetically dominated corona, when the thin disk is threaded by large-scale poloidal magnetic fields. This numerical result can naturally explain the coexistence of broaden Fe~\Ka line and hard X-ray continuum emission in XRBs and AGNs.

In this Letter, we consider the Zeeman effect on the Fe~\Ka lines of XRBs and AGNs in the MAD state assuming the two-pahse medium in the inner accretion disk. Future X-ray satellite missions such as X-Ray Imaging and Spectroscopy Mission ({\it XRISM}; \cite{XRISM2020SPIE11444E..22T}) and Advanced Telescope for High ENergy Astrophysics ({\it Athena}; \cite{Athena2013arXiv1306.2307N}) will carry an X-ray microcalorimeter with an energy resolution down to several~eV at 6~keV. We also discuss whether future X-ray missions can probe the MAD via the Zeeman effect.

\section{Zeeman Effect on MAD}

Magnetic flux $\Phi_\mathrm{BH}$ threading a BH is described as
\begin{equation}
    \Phi_\mathrm{BH}\equiv \phi_\mathrm{BH}(\dot{M}_\mathrm{BH}R_g^2c)^{1/2},
\end{equation}
where $\phi_\mathrm{BH}$ is the dimensionless magnetic flux, $\dot{M}_\mathrm{BH}$ is the accretion rate onto the BH, and $R_g=GM_\mathrm{BH}/c^2$ is the gravitational radius. $M_\mathrm{BH}$ is the BH mass, $G$ is the gravitational constant, and $c$ is the speed of the light. $\Phi_\mathrm{BH}$ in accretion flows are always nonzero since the magnetic flux is transported inward via accretion. $\phi_{\rm BH}$ is typically 20--50 depending on accretion rates based on GRMHD simulations (e.g., \cite{McKinney2012MNRAS.423.3083M, Avara2016MNRAS.462..636A, Liska2022ApJ...935L...1L}). We set $\phi_{\rm BH}=30$ as a fiducial value, which is based on the recent GR-radiation-MHD simulation accounting for both hot corona and cold medium at $\sim35$\% of the Eddington luminosity \citep{Liska2022ApJ...935L...1L}.

The MAD magnetic field at a distance $R\equiv r R_g $ from the BH is, assuming $BR^p=\mathrm{const.}$, 
\begin{eqnarray}
    B(R)&=&\frac{B(R_g)R_g^p}{R^p}\\
    &=&\frac{\Phi_\mathrm{BH}}{\pi R_g^2 r^p}\\
    &\simeq& 2.3\times10^9 \mathrm {G} \biggl(\frac{\phi}{30}\biggr)\biggl(\frac{m}{10}\biggr)^{-1/2}  \biggl(\frac{\dot{m}}{0.3}\biggr)^{1/2}\biggl(\frac{r}{1}\biggr)^{-p},
\end{eqnarray}
where $m\equiv M_{\mathrm BH}/M_\odot$ and $\dot{m}\equiv \dot{M}_\mathrm{BH}/\dot{M}_\mathrm{Edd}$ with a 10\% radiative efficiency. $\dot{M}_\mathrm{Edd}$ is the Eddington accretion rate for the mass of $M_\mathrm{BH}$. 

 The energy separation by the Zeeman effect is given by
\begin{equation}
     \Delta E_\mathrm{split} \approx \frac{e\hbar}{2m_ec} \left(M_L + 2M_S\right)B\simeq 11.6~\mathrm{eV}~ \biggl(\frac{B}{10^9~\mathrm{G}}\biggr) \label{eq:zeeman}
\end{equation}
where $M_L$ and $M_S$ are the quantum numbers of the orbital angular momentum and the spin angular momentum. We consider the split transition lines associated with changes of $\Delta M_L=\pm1$ and $\Delta M_S=0$. 

\begin{figure}[t]
\begin{center}
\includegraphics[width=\linewidth]{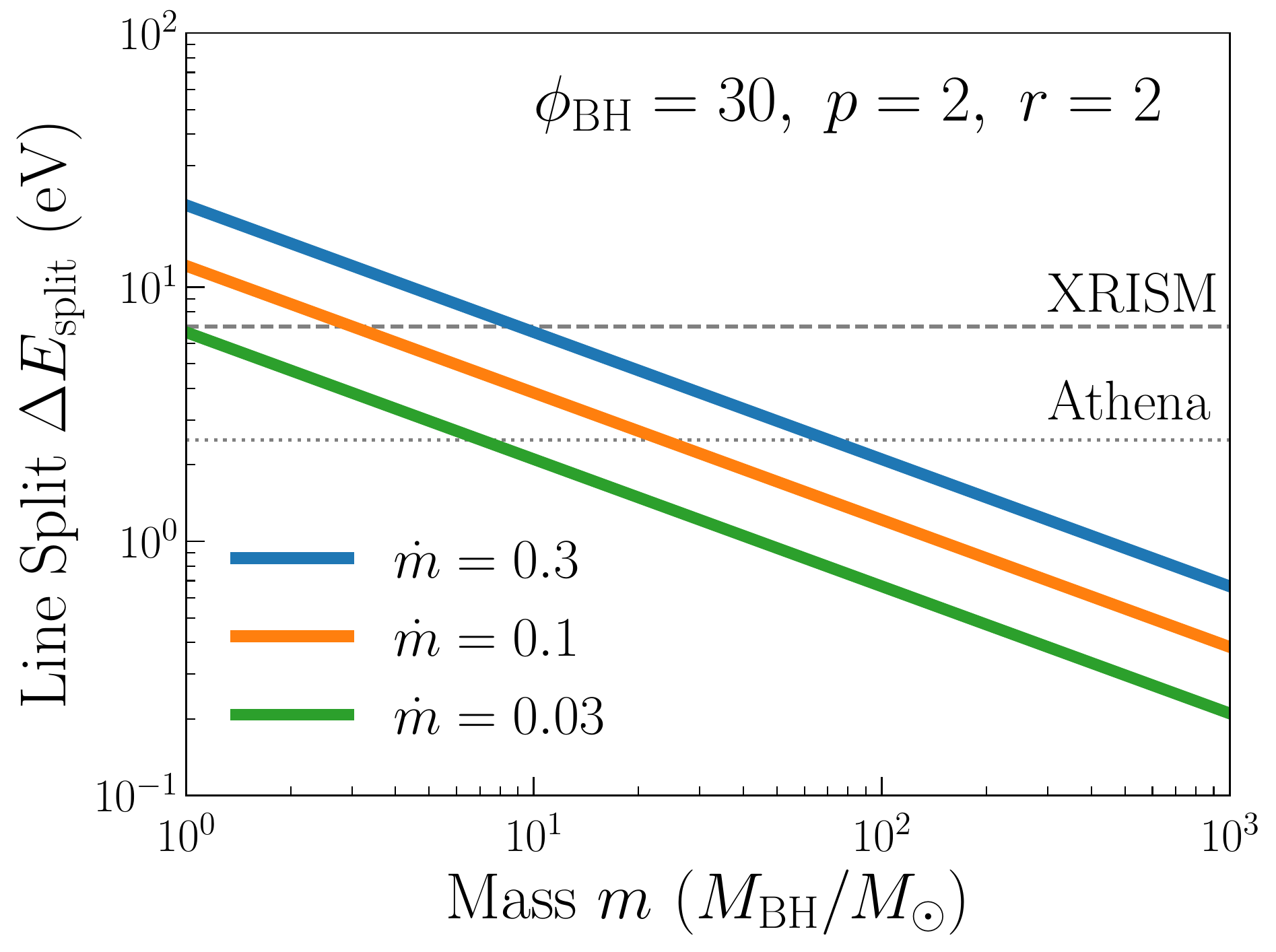}
\end{center} 
\caption{Fe~\Ka line splitting by the Zeeman Effect for various masses and accretion rates as indicated in the figure. We set $\phi_\mathrm{BH}=30$, $p=2$, and $r=2$. The horizontal dashed and dotted line represents the energy resolution of {\it XRISM} and {\it Athena}, respectively.
\label{fig:Zeeman}}
\end{figure}

In the accreting system, Eq.~\ref{eq:zeeman} can be replaced as
\begin{equation}
 \Delta E_\mathrm{split} \simeq 27~\mathrm{eV}~\biggl(\frac{\phi}{30}\biggr)\biggl(\frac{m}{10}\biggr)^{-1/2}  \biggl(\frac{\dot{m}}{0.3}\biggr)^{1/2}\biggl(\frac{r}{1}\biggr)^{-p}. \label{eq:split}
\end{equation}
Therefore, we can expect the line splitting of Fe~\Ka lines at the level of several tens of eV for XRBs and $10^{-3}$~eV for AGNs with $\phi_\mathrm{BH}=30$, $\dot{m}=0.3$, and $p=2$. 

The energy resolution of current X-ray CCDs at the Fe~\Ka line is $\sim120$~eV \citep{Ezoe2021RvMPP...5....4E}, larger than the typically expected split. With this resolution, we may infer $B<10^{10}$~G for any available observations. 
Next-generation X-ray telescopes such as {\it XRISM} and {\it Athena} will have X-ray micro-calorimeter instruments. The planned energy resolution is 7.0~eV and 2.5~eV at 6~keV for {\it XRISM} and {\it Athena}, respectively. Figure~\ref{fig:Zeeman} shows the expected Fe \Ka line split for various BH masses and accretion rates with $\phi_\mathrm{BH}=30$, $p=2$, and $r=2$. Figure~\ref{fig:Zeeman} also shows the energy resolutions of {\it XRISM} and {\it Athena}. Next-generation X-ray telescopes will enable us to see the Zeeman line splitting in XRBs with the presence of MAD.

About 20 XRBs have dynamically confirmed BHs \citep{Corral-Santana2016A&A...587A..61C}. Among them, low-mass black holes such as GX~339-4 and GRO~J1655-40, whose BH mass is $5.8\pm0.5~M_\odot$ \citep{Hynes2003ApJ...583L..95H} and $5.4\pm0.3~M_\odot$ \citep{Beer2002MNRAS.331..351B}, respectively, would be possible candidates to see the Zeeman effect on MAD. Another candidate is a persistent X-ray binary Cyg X-1 having the BH mass of $14.8\pm1.0~M_\odot$ \citep{Orosz2011ApJ...742...84O}\footnote{Recent radio astrometric observation suggests its mass of  $m=21.2\pm2.2$ \citep{Miller-Jones2021Sci...371.1046M}}. A broad iron \Ka line has also been reported for Cyg X-1 \citep{Duro2011A&A...533L...3D, Duro2016A&A...589A..14D}, with the inner edge of the disk of $\sim1.6R_g$ \citep{Fabian2012MNRAS.424..217F}. Although the accretion rate depends on the states, it is about 1\% even in the low/hard state \citep{Yamada2013PASJ...65...80Y}. The expected $\Delta E_\mathrm{split}$ is $2.7$ and $8.5$~eV for $\dot{m}=0.03$ and $0.3$, respectively. Therefore, {\it XRISM} and {\it Athena} would see the Zeeman effect on MAD even for a relatively massive stellar BH object Cyg~X-1.

\section{Discussions}

\begin{figure}[t]
\begin{center}
\includegraphics[width=\linewidth]{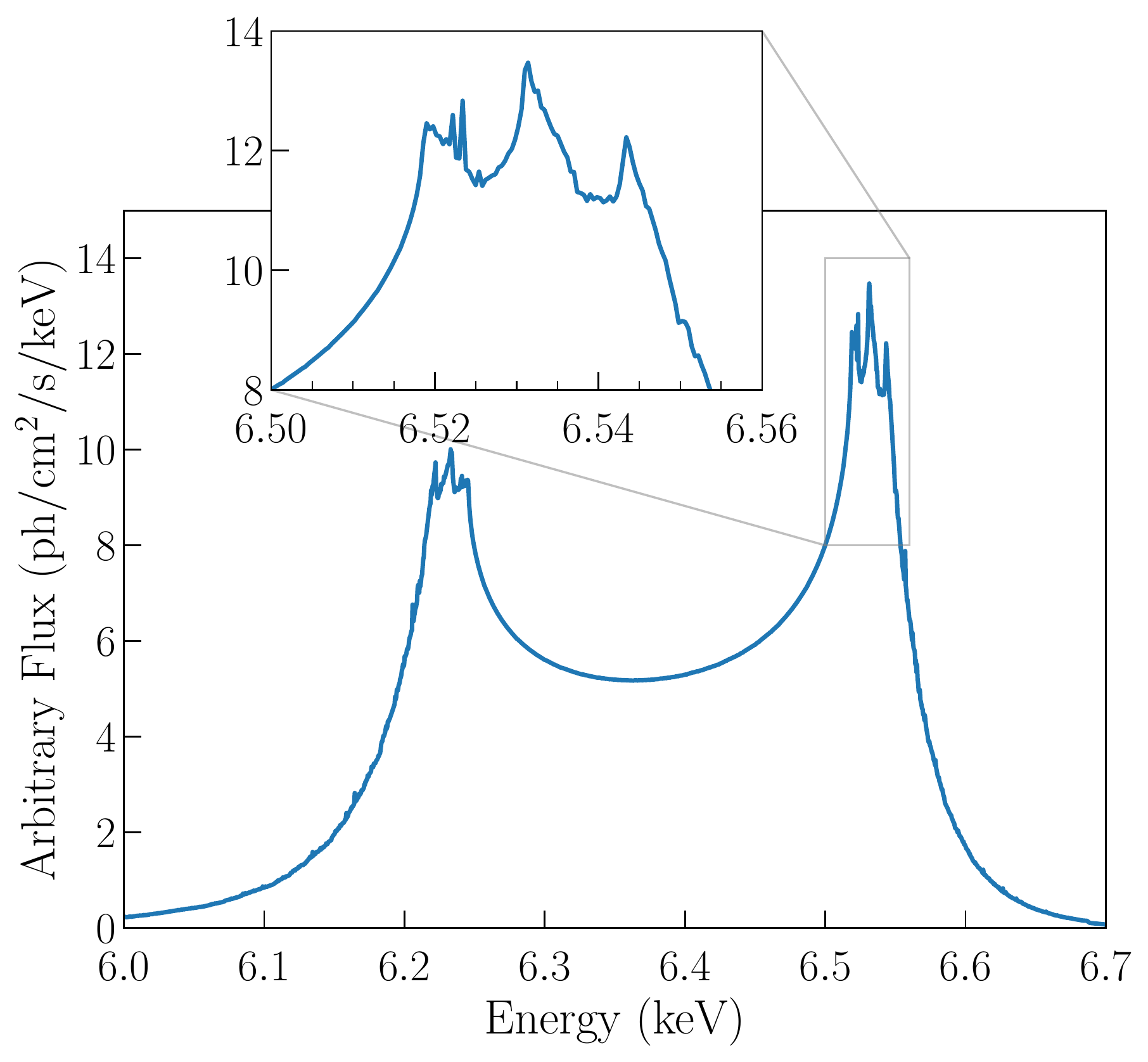}
\end{center} 
\caption{Simulated broad Fe~\Ka line spectrum accounting for the Zeeman Effect based on the {\tt kerrdisk} model. Detailed parameters are described in the text. We set $\Delta E_\mathrm{split}=24$~eV. The inset in the figure shows an enlarged view of the $6.50$ to $6.56$ keV region to clarify the split feature.
\label{fig:FeLine}}
\end{figure}

The Fe \Ka lines appear broadened and skewed by the Doppler effect and gravitational redshift \citep{Fabian1989MNRAS.238..729F, Laor1991ApJ...376...90L}. Those relativistic effects would blur the Zeeman split lines. Figure~\ref{fig:FeLine} shows a simulated Fe \Ka line in the MAD. We apply the {\tt kerrdisk} code \citep{Brenneman2006ApJ...652.1028B} with emissivity indices of $\alpha_1=\alpha_2=0$, inclination of $i=30^\circ$, dimensionless spin parameter of $a=0.9$, inner and outer radii of $r_\mathrm{min}=r_\mathrm{ms}$ and $r_\mathrm{max}=100~r_\mathrm{ms}$, and redshift $z=0$. $r_\mathrm{ms}$ is the marginally stable orbit radius. We include three lines split by the Zeeman effect assuming $\Delta E_\mathrm{split}=24~\mathrm{eV}$, corresponding to $B\simeq2\times10^{9}~\mathrm{G}$, we also set $p=0$. As the figure shows, three split lines would be distinguishable within $\Delta E_\mathrm{split}=24$~eV.  

There also exists Doppler broadening due to turbulent motions of accreting flows. The turbulence speed is characterized by the sound speed. The broadening by turbulence is expected at the level of $\sim10$~eV at the temperature of $\sim10^7$~K, which can be comparable to the expected Zeeman split (Eq.~\ref{eq:split}). Three split lines would be more broaden by turbulence than shown in Figure~\ref{fig:FeLine}. In addition, the effect of the continuum is not included in Figure~\ref{fig:FeLine} and the energy split would depend on radius if $p\neq0$. Further detailed spectral simulations, including instrument response functions of {\it XRISM} and {\it Athena}, will be needed.

Broad Fe \Ka lines are commonly reported in various XRBs and AGNs \citep{Reynolds2021ARA&A..59..117R}. However, line broadening is known to depend on the spectral modeling (see, e.g., \cite{Done1999MNRAS.305..457D, Makishima2008PASJ...60..585M}) and disk winds further would disturb the ionization states (see e.g., \cite{Tomaru2019MNRAS.490.3098T}). If the location of reflecting iron atoms is further away from the BH, the expected line split drops with $r^{-p}$. Determination of the reflecting medium is necessary to probe the MAD through X-ray Zeeman effect measurements. Therefore, the Zeeman effect on the Fe \Ka line is realized only with the MAD state and the near BH reflector case. In other words, if we do not see the Zeeman effect even with sufficient energy resolutions, it will imply that MAD is absent or the reflector is distant. 

With a strong magnetic field like MAD, we would also have the quadratic Zeeman effect producing displacements of lines toward shorter wavelengths \citep{Jenkins1939PhRv...55...52J, Schiff1939PhRv...55...59S, Preston1970ApJ...160L.143P, Loeb2003PhRvL..91g1103L}. The energy shift is given as
\begin{eqnarray}
&&\Delta E_\mathrm{shift} = \frac{e^2a_0^2}{8Z^2m_ec^2}n^4(1+M_L^2)B^2\\
&&\simeq 9.6\times10^{-3}~\mathrm{eV}~\biggl(\frac{\phi}{30}\biggr)^2\biggl(\frac{m}{10}\biggr)^{-1}  \biggl(\frac{\dot{m}}{0.3}\biggr)\biggl(\frac{r}{1}\biggr)^{-2p} \label{eq:shift},
\end{eqnarray}
where $n$ is the principal quantum number, $a_0$ is  the Bohr radius, and $Z$ is the nuclear charge. We set $n=1$, $M_L=1$ and $Z=26$ in Eq.~\ref{eq:shift}. Thus, several orders of magnitude better energy resolution would be necessary to see the quadratic energy shift.

\section{Summary}
In this Letter, we consider the Zeeman effect on the MAD state. MAD is expected to be associated with the jet production \citep{Tchekhovskoy2010ApJ...711...50T, Tchekhovskoy2011MNRAS.418L..79T, McKinney2012MNRAS.423.3083M}. In the black hole accretion systems, broad Fe~\Ka fluorescence lines have been often reported. A strong magnetic field environment by MAD would induce line splitting of the Fe~\Ka line by the Zeeman effect (Eq.~\ref{eq:split}). Next-generation X-ray telescopes such as {\it XRISM} and {\it Athena} will be able to see the Zeeman splitting of Fe lines in XRBs, if reflectors exist near BHs. The detection of the Zeeman effect would be clear evidence of the MAD in the BH accretion systems. If the Zeeman effect does not appear even with sufficient energy resolutions, it would imply that MAD is absent or the iron reflector is distant.

\begin{ack}
We would like to thank the anonymous referee for thoughtful and helpful comments. We would also like to thank Roger Blandford, Chris Done, Norita Kawanaka, Katsunori Kusakabe, Shin Mineshige, Hirokazu Odaka,and Shinsuke Takasao for useful discussions and comments. Y.I. is supported by JSPS KAKENHI Grant Number JP18H05458, JP19K14772, and JP22K18277. This work was supported by World Premier International Research Center Initiative (WPI), MEXT, Japan.  
\end{ack}

\end{document}